\pdfoutput=1
\documentclass[english,manuscript]{aastex}
\usepackage[T1]{fontenc}
\usepackage[latin9]{inputenc}
\usepackage{amsbsy}
\usepackage{amssymb}
\usepackage{graphicx}
\usepackage{esint}

\makeatletter

\providecommand{\tabularnewline}{\\}

\shorttitle{Surface-shear shaped solar dynamo}
\shortauthors{Pipin \& Kosovichev}
\bibpunct[,]{(}{)}{;}{a}{,}{,}

\usepackage{babel}

\usepackage{babel}

\usepackage{babel}

\usepackage{babel}

\makeatother

\usepackage{babel}
\begin{document}

\title{Mean-field solar dynamo models with strong meridional flow at the
bottom of the convection zone}

\author{V.V. Pipin$^{1-3}$ and A.G. Kosovichev$^{3}$}

\affil{ $^{1}$ Institute of Geophysics and Planetary Physics, UCLA, Los
Angeles, CA 90065, USA \\
 $^{2}$Institute of Solar-Terrestrial Physics, Russian Academy of
Sciences, \\
 $^{3}$Hansen Experimental Physics Laboratory, Stanford University,
Stanford, CA 94305, USA }
\begin{abstract}
The paper presents a study of kinematic axisymmetric mean-field dynamo
models for a case of the meridional circulation with a deep-seated
stagnation point and a strong return flow at the bottom of the convection
zone. This kind of circulation follows from mean-field models of the
angular momentum balance in the solar convection zone. We show that
it is possible for this types of meridional circulation to construct
kinematic dynamo models that resemble in some aspects the sunspot
magnetic activity cycle. The dynamo model includes turbulent sources
of the large-scale poloidal magnetic field production, due to kinetic
helicity and a combined effect due to Coriolis force and the large-scale
current. In these models the toroidal magnetic field, which is responsible
for the sunspot production, is concentrated at the bottom of the convection
zone, and is transported to low-latitude regions by the meridional
flow. The meridional component of the poloidal field is also concentrated
at the bottom of the convection zone while the radial component is
concentrated in near polar regions.
There are some issues which, perhaps, are resulted from the given
meridional circulation pattern and the distribution of the magnetic
diffusivity inside convection zone. In particular, in the near-equatorial
regions the phase relations between the toroidal and poloidal components
disagree with observations. Also, we show that the period of the magnetic
cycle may not always monotonically decrease with the increase of the
meridional flow speed. Thus, for the further progress it is important
to determine the structure of the meridional circulation, which is
one of the critical properties, from helioseismology observations. 
\end{abstract}

\section{Introduction}

The widely accepted paradigm about the nature of the global magnetic
activity of the Sun assumes that meridional circulation is an important
part of the dynamo processes operating in the solar convection zone
\citep{1995A&A...303L..29C,1995SoPh..160..213D,1999SoPh..184...61C,2010arXiv1009.6184M}.
The current flux-transport and mean-field dynamo models, (e.g., \citealp{1999ApJ...518..508D,2004MNRAS.350..317G,2002A&A...390..673B}),
typically employ an analytical profile of the meridional circulation
pattern, which has to satisfy a mass conservation equation and the
relevant boundary conditions. One of the basic features of this profile
is that the circulation stagnation point is close to the middle of
the convection zone, (e.g., \citealp{2002A&A...390..673B,2004ApJ...601.1136D}).
 However, such ad hoc models of the meridional flow have no support
from the mean-field theory of the angular momentum distribution in
the solar convection zone \citep{1999A&A...344..911K,2011MNRAS.411.1059K},
also see, \citet{2009ApJ...704....1G}. The theory predicts the meridional
circulation pattern with nearly equal amplitudes of the flow velocity
at the bottom and top of the convection zone. The stagnation point
of this flow is close to the bottom of convection zone, near $0.75R_{\odot}$,
and the circulation is concentrated near the convection zone boundaries.
Rempel\citeyearpar{rem2005ApJ,2006ApJ...647..662R} obtained a similar
meridional circulation profile with a deep stagnation point and used it to construct a nonlinear
dynamo model. 

The physical mechanisms of the strong deviation of the meridional
circulation pattern from the simple analytical models are discussed
in the recent paper by \citet{2011MNRAS.411.1059K}. Here, we briefly
summarize their main arguments. The distribution of large-scale flows
in the balk of convection zone is close to the Taylor-Proudman balance.
However, this balance is violated near the boundaries. This results
in a concentration of the velocity circulation in the Eckman layers
near the bottom and the top of convection zone \citep{1999ApJ...511..945D,2006ApJ...641..618M,2010A&A...510A..33B}.

Our goal is to investigate how the meridional circulation with a fast
return flow at the bottom of the convection zone can affect solar
dynamo models. We construct a series of kinematic mean-field dynamo
models that employ the meridional circulation pattern suggested
by \citet{2011MNRAS.411.1059K}. These dynamo models include the turbulent
generation of the magnetic field due to the kinetic helicity ( $\alpha$-effect),
the combined effect of the Coriolis force and large-scale current
($\Omega\times J$-effect), and the toroidal magnetic field generation
due to the differential rotation ($\Omega$-effect). Following Krause
and Rädler(1980) these models can be classified as $\alpha^{2}\delta\Omega$
dynamo. Our approach is to investigate conditions of the dynamo instability
for this type of the meridional circulation and determine the basic
properties of the dynamo solution at the instability threshold. This
is a kinematic dynamo problem. The next section describes the formulation
of the mean-field dynamo model, including the basic assumptions, the
reference model of the solar convection zone, and input parameters
of the large-scale flows. Section 3 presents the results and discussion.
The main findings are summarized in Section 4.

\section{Basic equations}

\subsection{Formulation of model}

The dynamo model is based on the standard mean-field induction equation
in perfect conductive media (Krause and Rädler, 1980): 
\begin{equation}
\frac{\partial\mathbf{B}}{\partial t}=\boldsymbol{\nabla}\times\left(\mathbf{\boldsymbol{\mathcal{E}}+}\mathbf{U}\times\mathbf{B}\right),\label{induc}
\end{equation}
 where $\boldsymbol{\mathcal{E}}=\overline{\mathbf{u\times b}}$ is
the mean electromotive force, with $\mathbf{u,\, b}$ being the turbulent
fluctuating velocity and magnetic field respectively; $\mathbf{U}$
is the mean velocity. General expression for $\boldsymbol{\mathcal{E}}$
was computed by \citet{2008GApFD.102...21P}(hereafter P08). Following
Krause and Rädler(1980) we write the expression for the mean-electromotive
force as follows: 
\begin{equation}
\mathcal{E}_{i}=\left(\alpha_{ij}+\gamma_{ij}\right)\overline{B}-\eta_{ijk}\nabla_{j}\overline{B}_{k}\label{eq:EMF}
\end{equation}
 where, tensor $\alpha_{i,j}$ represents the turbulent alpha effect,
tensor $\gamma_{i,j}$ describes the turbulent pumping, and the $\eta_{ijk}$
term describes the anisotropic diffusion due to the Coriolis force
and the $\Omega\times J$ effect \citep{rad69}.

We consider a large-scale axisymmetric magnetic field, $\mathbf{B}=\mathbf{e}_{\phi}B+\nabla\times\frac{A\mathbf{e}_{\phi}}{r\sin\theta}$,
where $B(r,\theta,t)$ is the azimuthal component of the magnetic
field, $A(r,\theta,t)$ is proportional to the azimuthal component
of the vector potential, $r$ is radial coordinate and $\theta$ -
polar angle. The mean flow is given by velocity vector $\mathbf{U}=\mathbf{e}_{r}U_{r}+\mathbf{e}_{\theta}U_{\theta}+\mathbf{e}_{\phi}r\sin\theta\Omega$,
where $\Omega\left(r,\theta\right)$ is the angular velocity of the
solar differential rotation and $U_{r}(r,\theta)$, $U_{\theta}(r,\theta)$,
represent velocity components of the meridional circulation. The mean-field
magnetic field evolution of is governed by the dynamo equations, which
follow from Eq.(\ref{induc}): 
\begin{eqnarray}
\frac{\partial A}{\partial t} & = & r\sin\theta\mathcal{E}_{\phi}+\frac{U_{\theta}\sin\theta}{r}\frac{\partial A}{\partial\mu}-U_{r}\frac{\partial A}{\partial r}\label{eq:A}\\
\frac{\partial B}{\partial t} & = & -\frac{\sin\theta}{r}\left(\frac{\partial\Omega}{\partial r}\frac{\partial A}{\partial\mu}-\frac{\partial\Omega}{\partial\mu}\frac{\partial A}{\partial r}\right)-\frac{\partial\left(rU_{r}B\right)}{\partial r}+\frac{\sin\theta}{r}\frac{\partial U_{\theta}B}{\partial\mu}+\frac{1}{r}\frac{\partial r\mathcal{E}_{\theta}}{\partial r}+\frac{\sin\theta}{r}\frac{\partial\mathcal{E}_{r}}{\partial\mu}\label{eq:B}
\end{eqnarray}
 We introduce the free parameter $C_{\eta}$ to control the turbulent
diffusion coefficient (see Appendix). A solar-type dynamo model can
be constructed not only with the $\alpha$-effect as a prime turbulent
source of the poloidal magnetic field generation \citep{stix76a,2009A&A...493..819P,2009A&A...508....9S}.
The exact mechanism of the large-scale poloidal magnetic field production
on the Sun is not known. After \citet{1955ApJ...122..293P}, it is
belived that the $\alpha$-effect (associated with cyclonic convection)
is the most important turbulent source of the poloidal magnetic field
generation on the Sun. The $\alpha$-effect particularly important role
for the dynamo because it describes a leading term of the Taylor
expansion of the turbulent mean-electromotive force in terms of the scale separation
parameter. 
In addition, the mean-field theory predicts the magnetic field generation
effects  due to the interaction of the Coriolis force
($\Omega\times J$-effect) and the differential rotation ($W\times J$-effect,
where $W$ is the large-scale velocity shear) with large-scale electric
current (see, \citealp{rad69,krarad80,kle-rog:04a}). The role of
these mechanisms for the solar dynamo is not well understood though
it was shown (\citealp{2008arXiv0812.1466P,2009A&A...493..819P})
that it is possible to construct the solar-type dynamo models including
these effects. Our model includes the $\alpha$- and $\Omega\times J$
effects.  We introduce free parameters $C_{\alpha}$
and $C_{\delta}^{(\Omega)}$ to control their strength.

We use the solar convection zone model computed by \citet{stix:02},
in which mixing length $\ell=\alpha_{MLT}\left|\Lambda^{(p)}\right|^{-1}$,
where $\mathbf{\boldsymbol{\Lambda}}^{(p)}=\boldsymbol{\nabla}\log\overline{p}$
is the inverse pressure scale height, and $\alpha_{MLT}=2$. 
 We confine the integration domain between 0.712$R_{\odot}$
and 0.972$R_{\odot}$ in radius. It extends from the pole to pole
in latitude. The differential rotation profile, $\Omega=\Omega_{0}f_{\Omega}\left(x,\mu\right)$,
$x=r/R_{\odot}$, $\mu=\cos\theta$ is a modified version of the analytical
approximation to helioseismology data proposed by \citet{1998MNRAS.298..543A},
see Figure 1a.

\begin{figure}
\begin{centering}
a)\includegraphics[width=0.28\textwidth]{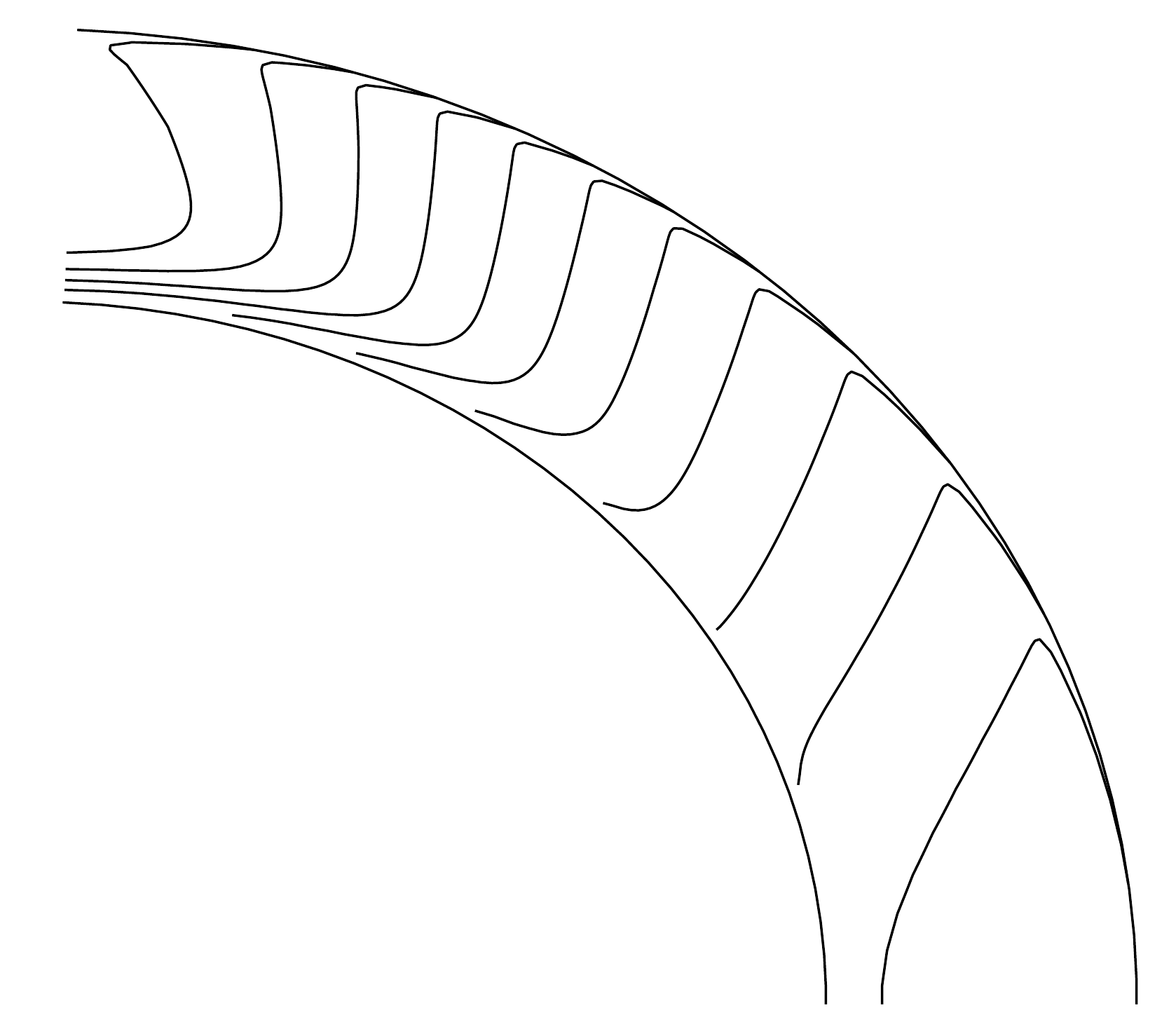}b)\includegraphics[width=0.28\textwidth]{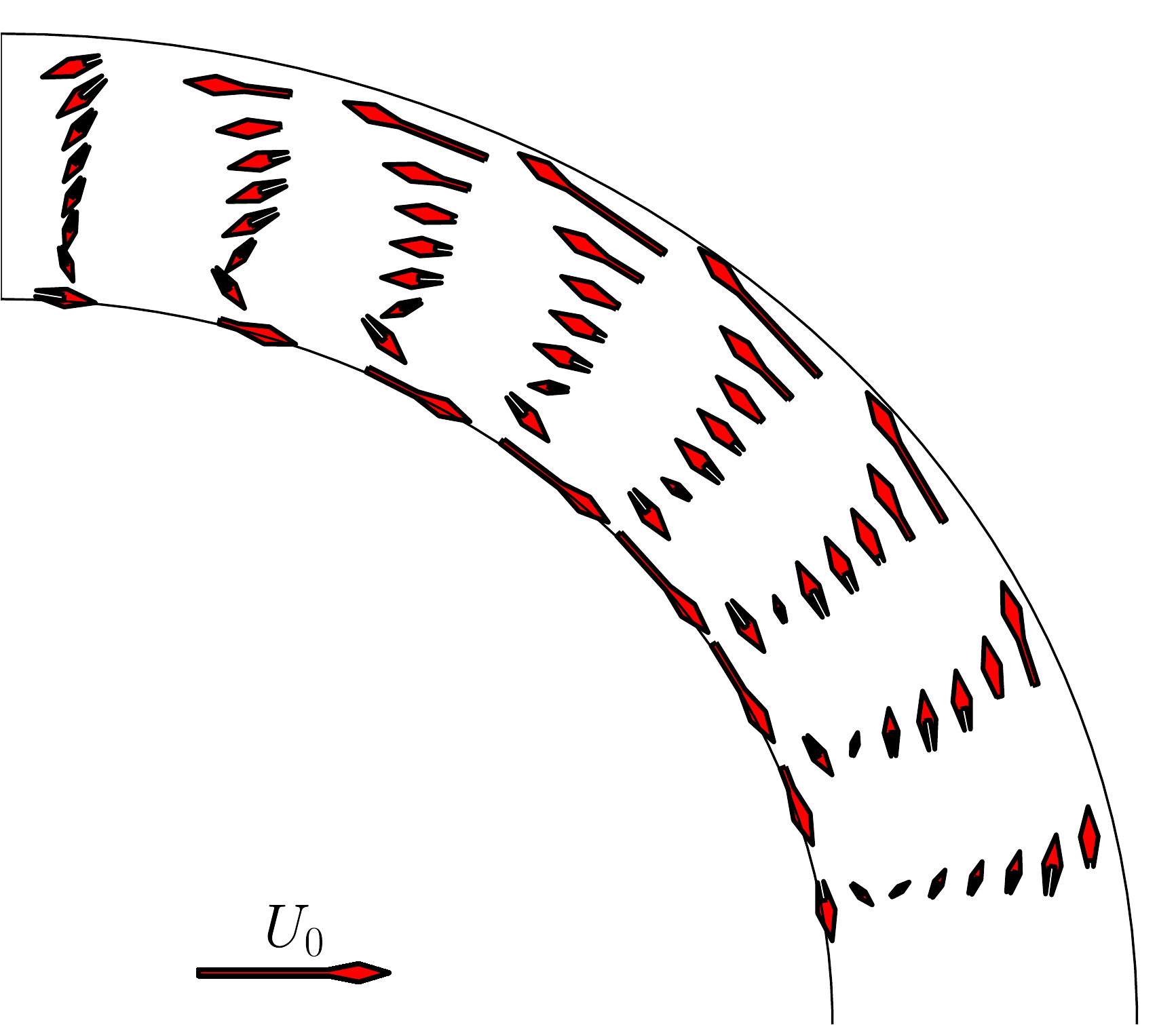}c)
\includegraphics[width=0.35\textwidth]{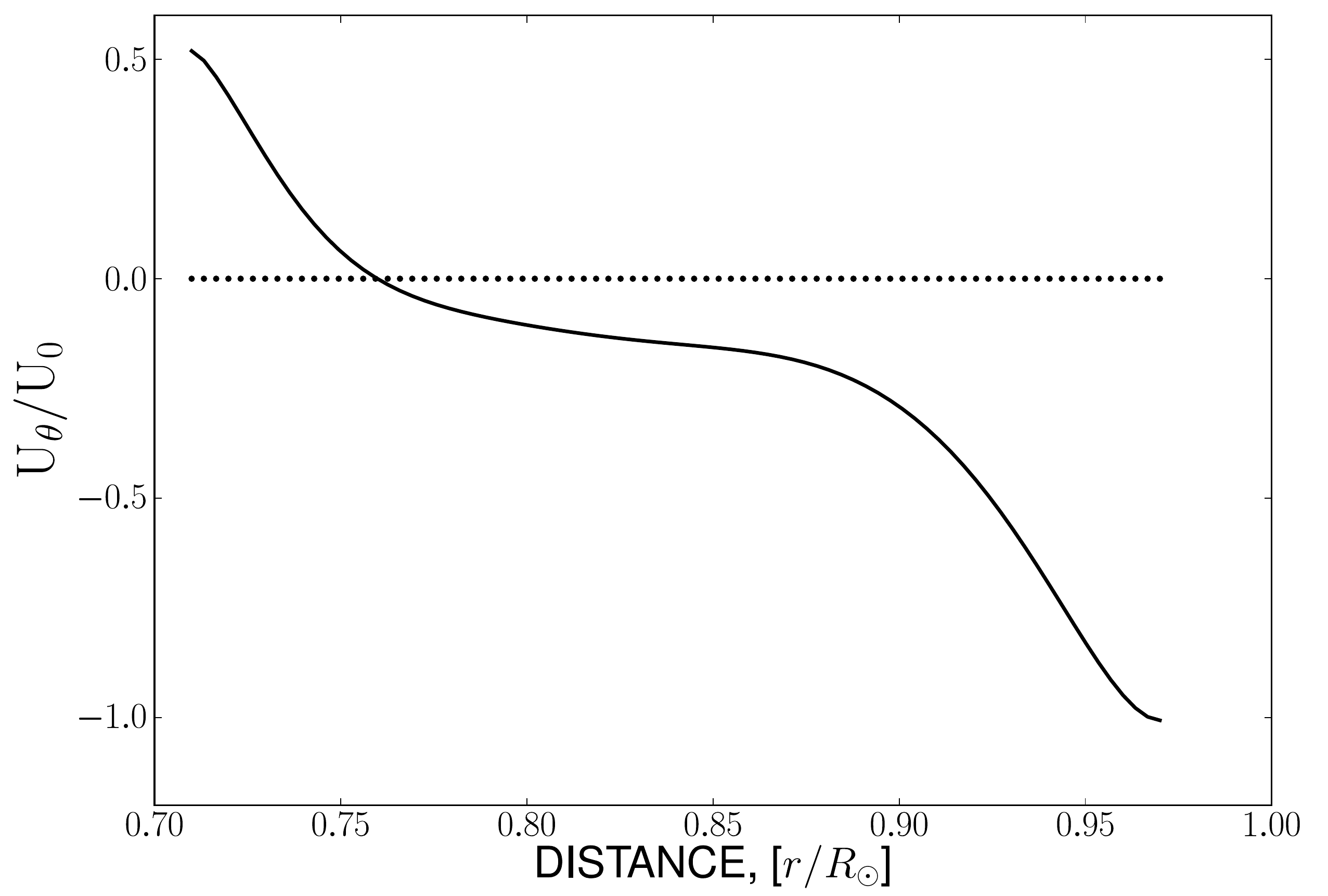} 
\par\end{centering}

\begin{centering}
d)\includegraphics[width=0.5\textwidth]{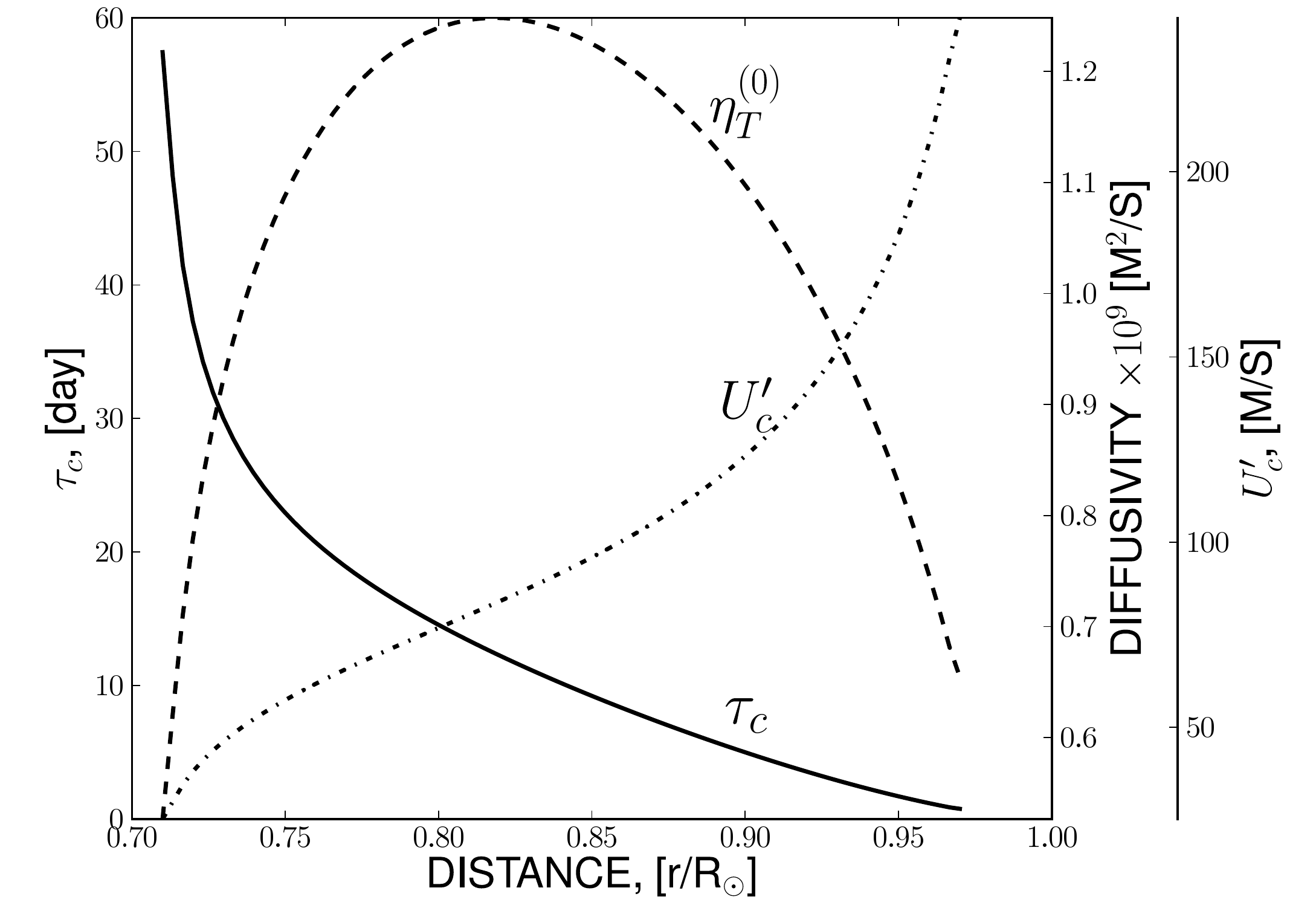}e)\includegraphics[width=0.4\textwidth]{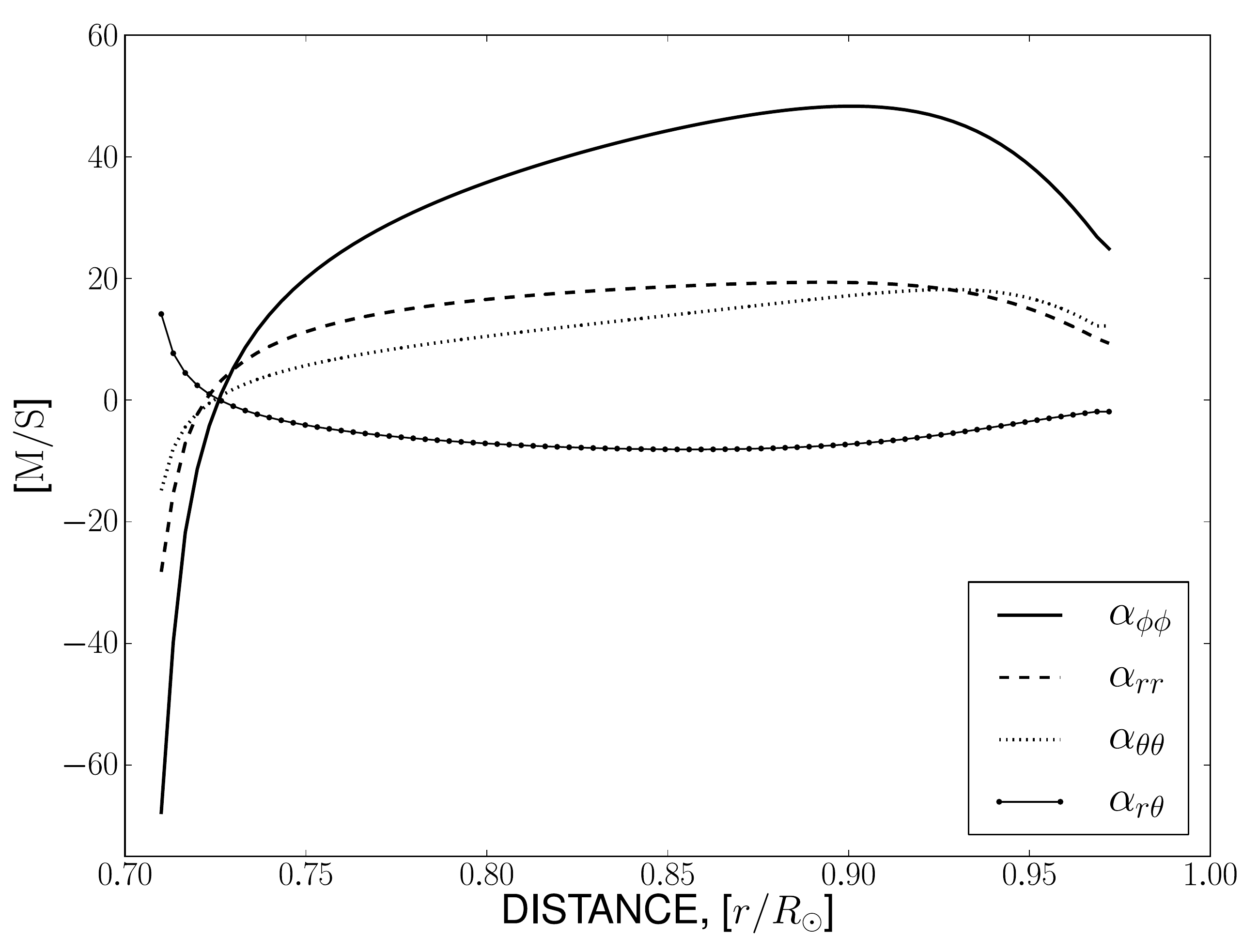} 
\par\end{centering}

\caption{\label{fig1}Internal parameters of the solar convection zone: a)
the contours of the constant angular velocity are plotted for the
levels $(0.75-1.05)\Omega_{0}$ with a step of $0.025\Omega_{0}$,
$\Omega_{0}=2.86\cdot10^{-7}s^{-1}$; b) the vector field of the meridional
circulation with $\left|\mathbf{U}\right|$ measured in units of $U_{0}$.
c)the meridional component of circulation at $\theta=45^{\circ}$;
d) turnover convection time $\tau_{c}$, the background turbulent
diffusivity $\eta_{T}^{(0)}$, RMS convective velocity $U'_{c}$;
e) the radial profiles of the $\alpha$-effect components at $\theta=45^{\circ}$;}
\end{figure}

The meridional flow is modeled in the form of two stationary circulation
cells, one in the northern and one in the southern hemisphere, with
a poleward motion in the upper and equator-ward motion in the lower
part of the convection zone. Following \citet{2011MNRAS.411.1059K},
the meridional circulation velocity components were approximated via
the orthogonal Chebyshev polynomial decompositions: 
\begin{eqnarray}
U_{\theta} & = & 3U_{0}\sin\theta\cos\theta\sum_{n=0}^{3}c_{\theta}^{(n)}T_{n}\left(\xi\right),\label{eq:kit-t}\\
U_{r} & = & U_{0}\left(3\cos^{2}\theta-1\right)\left(1-\xi^{2}\right)\sum_{n=0}^{3}c_{r}^{(n)}T_{n}\left(\xi\right),\label{eq:kit-r}
\end{eqnarray}
 where 
\begin{equation}
\xi=\frac{2x-x_{e}-x_{b}}{x_{e}-x_{b}}.\label{eq:tran}
\end{equation}
 Here, $x_{b,e}$ are the radial boundaries of the integration domain.
In our case, $x_{b}=0.712$ and $x_{e}=0.972$. The coefficients $c_{\theta}^{(n)}$,
$c_{r}^{(n)}$ are given in Table \ref{tabl1}. Parameter $U_{0}$
controls the speed of the meridional circulation. \citet{2011MNRAS.411.1059K}
obtain $U_{0}\approx16$ $\mathrm{ms^{-1}}$.

\begin{table}
\begin{centering}
\begin{tabular}{|c|c|c|c|c|}
\hline 
n  & 0  & 1  & 2  & 3\tabularnewline
\hline 
\hline 
$c_{\theta}^{(n)}$  & -0.13432(5)  & - 0.40473(6)  & - 0.02170(3)  & - 0.10718(5)\tabularnewline
\hline 
$c_{r}^{(n)}$  & - 0.0681469(4)  & - 0.006839(4)  & -0.032516(1)  & - 0.0027(4)\tabularnewline
\hline 
\end{tabular}
\par\end{centering}

\caption{\label{tabl1}The coefficients for  the
meridional circulation profile components given by Eqs.(\ref{eq:kit-t},\ref{eq:kit-r}).}
\end{table}

The geometry of the meridional flow is illustrated in Figure \ref{fig1}(b).
Figure \ref{fig1}(c) shows the latitudinal component of the circulation
in units $U_{0}$ .

The boundary conditions represent a perfect conductor at the bottom
and the potential magnetic field configuration outside the domain.

\subsection{Method of solution}

We investigate the linear stability of the dynamo equations (\ref{eq:A},\ref{eq:B})
and determine unstable dynamo modes. Then we construct linear dynamo
solutions using the corresponding eigen-functions. Our approach to
solve the linear problem was described in details by \citet{2009A&A...493..819P}
and \citet{2009A&A...508....9S}. We use a Galerkin method, expanding
the magnetic field in terms of a basis that satisfies the boundary
conditions \citep{boyd,2005GApFD..99..467L}. The system of Eqs.~(\ref{eq:A})
and (\ref{eq:B}) has exponentially growing or decaying solutions,
which we represent in the form 
\begin{eqnarray}
A\left(x,\theta,t\right) & = & \mathrm{e}^{{\displaystyle \sigma t}}\sum_{n}\sum_{m}A_{nm}\sin\theta\, S_{nm}^{(A)}\left(\xi\right)P_{m}^{1}\left(\cos\theta\right),\label{eq:Am}\\
B\left(x,\theta,t\right) & = & \mathrm{e}^{{\displaystyle \sigma t}}\sum_{n}\sum_{m}B_{nm}S_{n}^{(B)}\left(\xi\right)P_{m}^{1}\left(\cos\theta\right),\label{eq:Bm}
\end{eqnarray}
 where $S_{nm}^{(A)}\left(\xi\right)$ and $S_{n}^{(B)}\left(\xi\right)$
are linear combinations of Legendre polynomials, and $P_{m}^{1}$
is the associated Legendre function of degree $m$ and order $1$.
These expansions ensure the regularity of the solutions at the poles
$\theta=0$ and $\theta=\pi$. The integrations in radius and latitude,
which are needed for calculating the expansion coefficients $A_{nm}$
and $B_{nm}$, were done by means of the Gauss-Legendre procedure.
The eigenvalue problem for determining the eigenvalues, $\sigma$,
and the associated eigenfunctions was solved by using the LAPACK software.
There are two types of dynamo eigenmodes: 1) modes with a symmetric
distribution of the toroidal component $B$ and antisymmetric distribution
of the poloidal component $A$, relative to the equator, called here
{}``S-modes'', and 2) vice-versa antisymmetric al modes, called
{}``A-modes''. We define the eigenvalues of the S- and A-modes as
$\sigma^{(S)}=\lambda^{(S)}+i\omega^{(S)}$ and $\sigma^{(A)}=\lambda^{(A)}+i\omega^{(A)}$.
The spectral resolution of our calculations was 16 radial and 25 latitudinal
basis functions. The results were qualitatively confirmed by a number
of runs with larger number of the basis functions.

\begin{figure}
\includegraphics[width=0.4\paperwidth,height=0.2\paperheight]{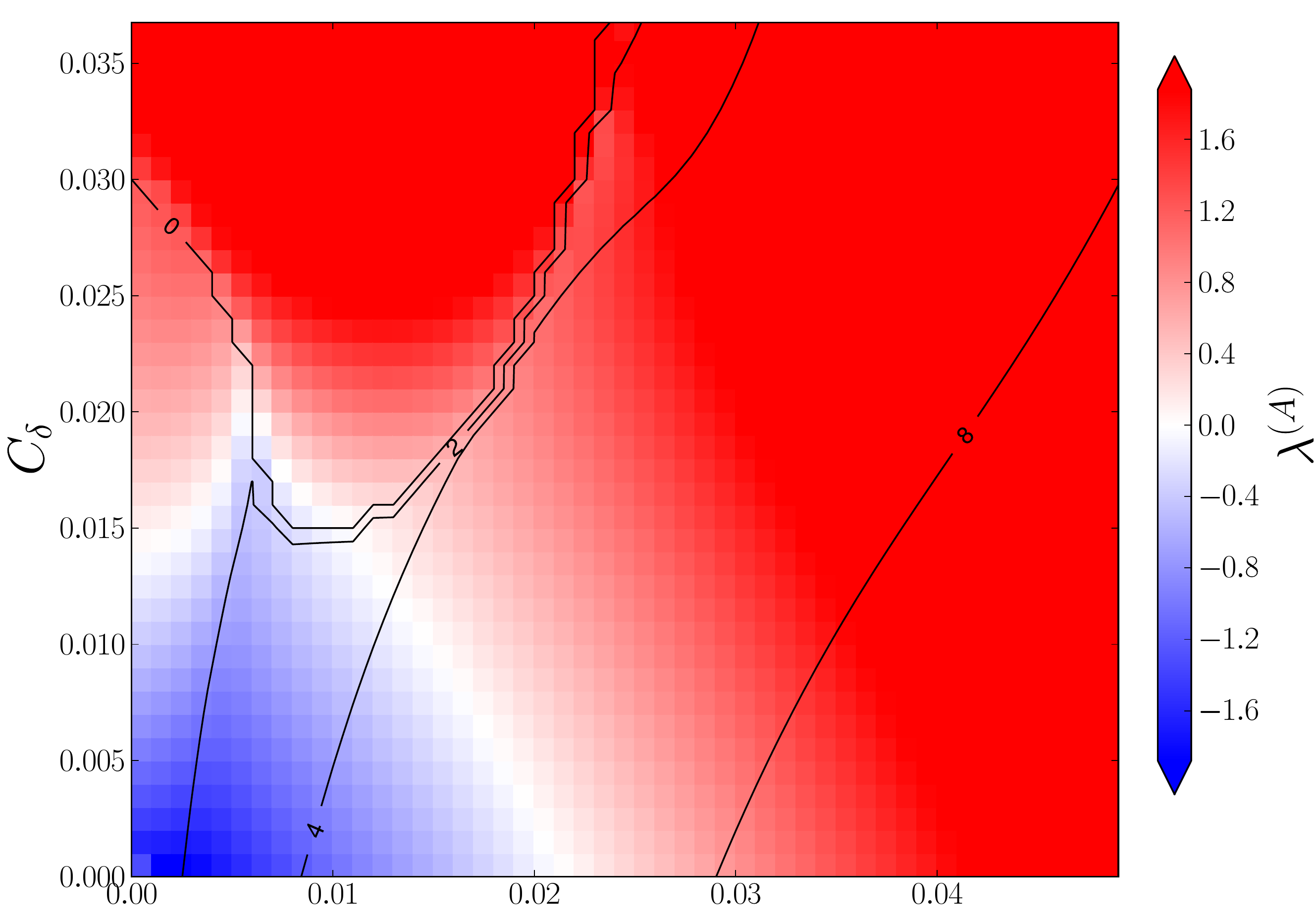}\includegraphics[width=0.4\paperwidth,height=0.2\paperheight]{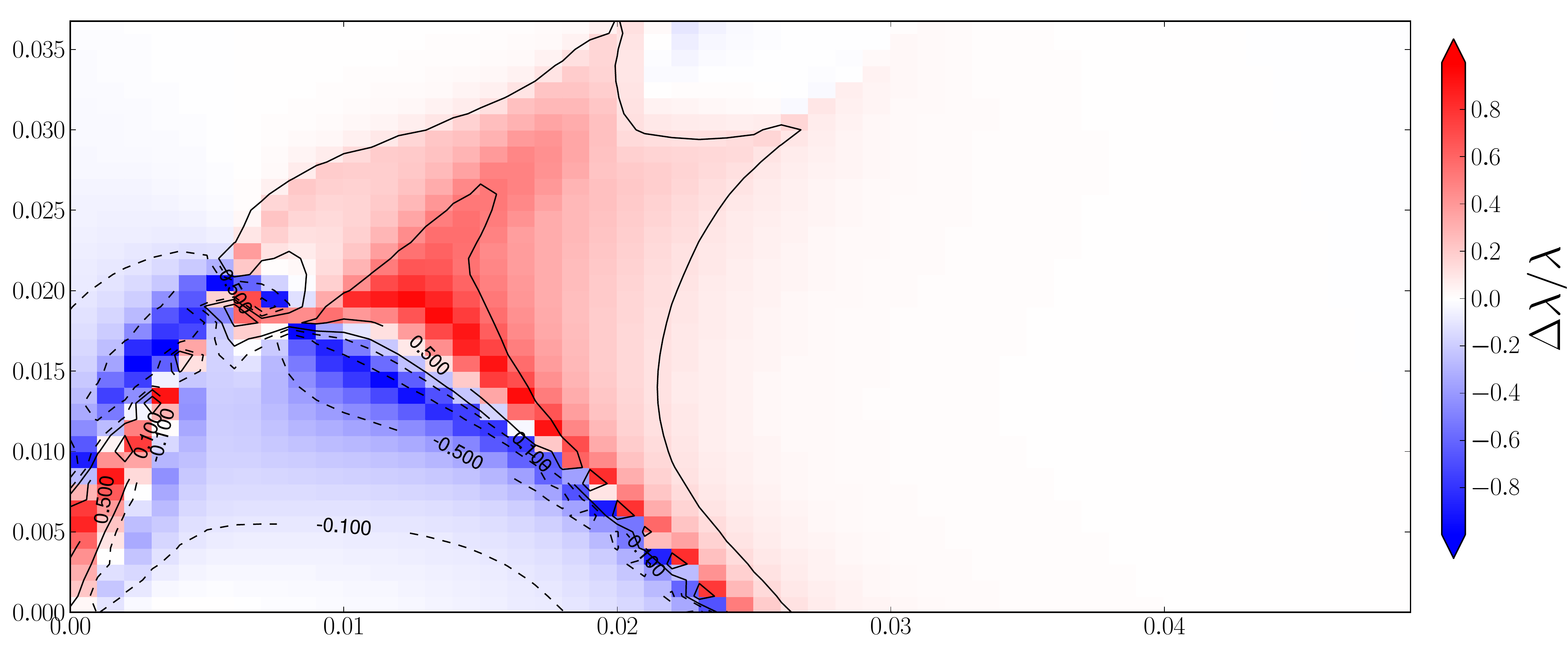}

\includegraphics[width=0.4\paperwidth,height=0.2\paperheight]{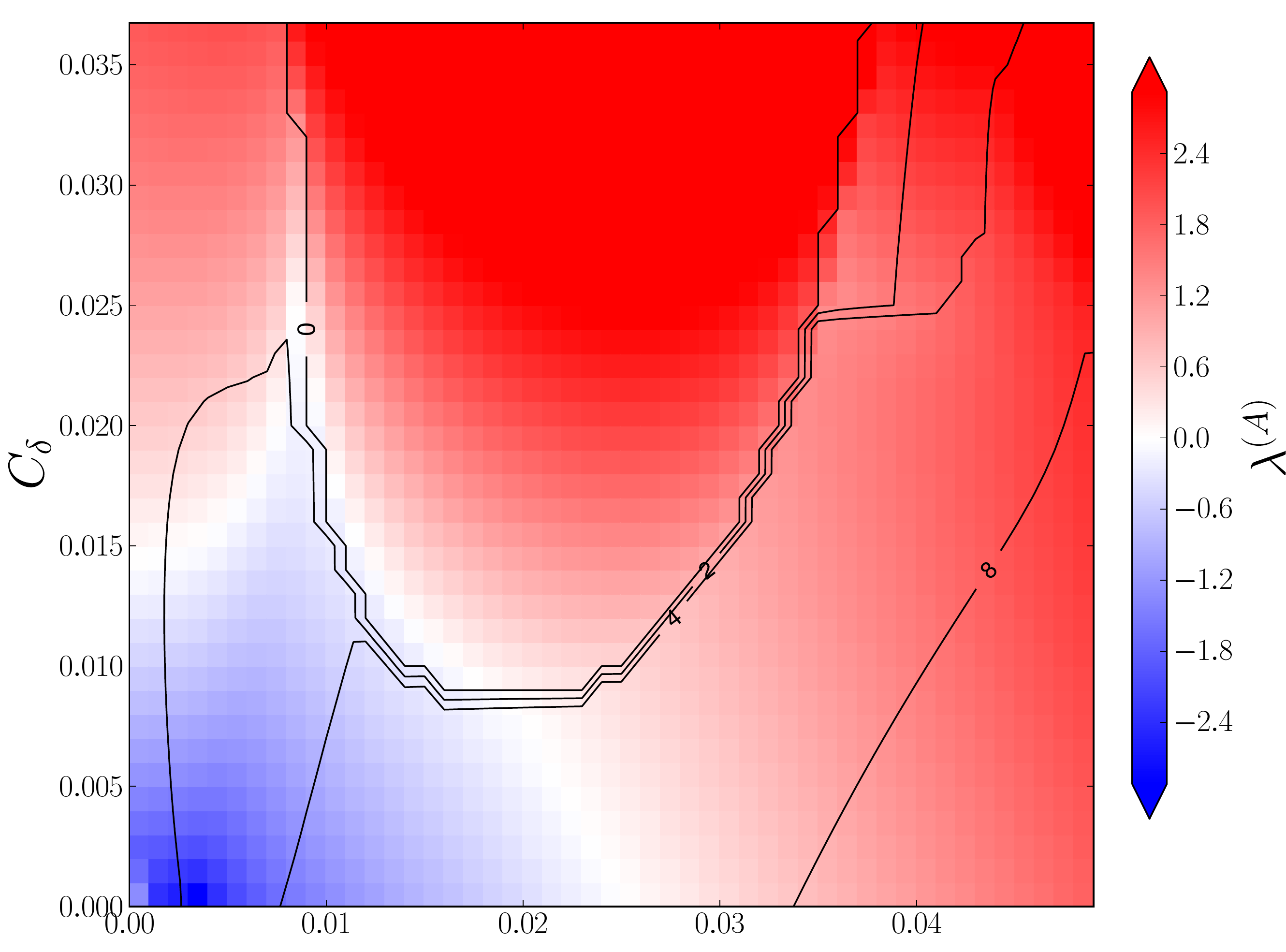}\includegraphics[width=0.4\paperwidth,height=0.2\paperheight]{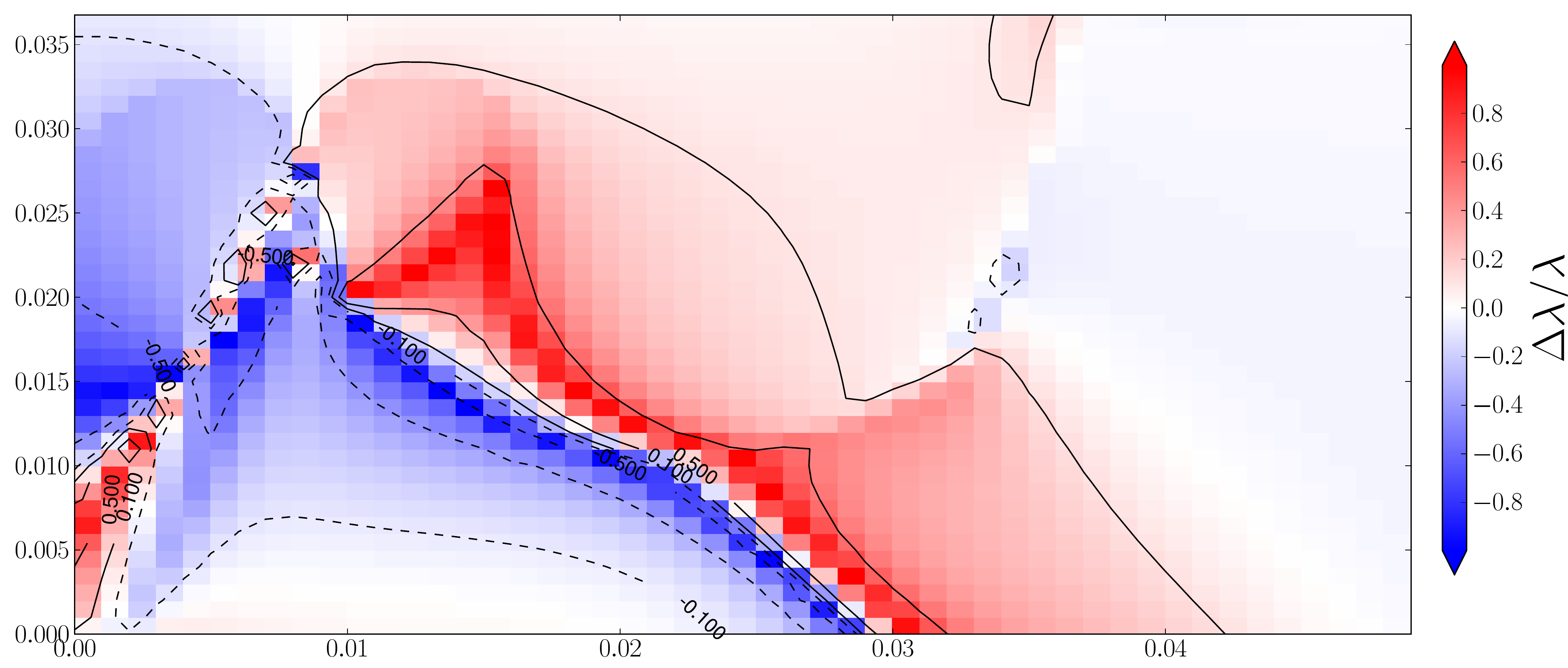}

\includegraphics[width=0.4\paperwidth,height=0.2\paperheight]{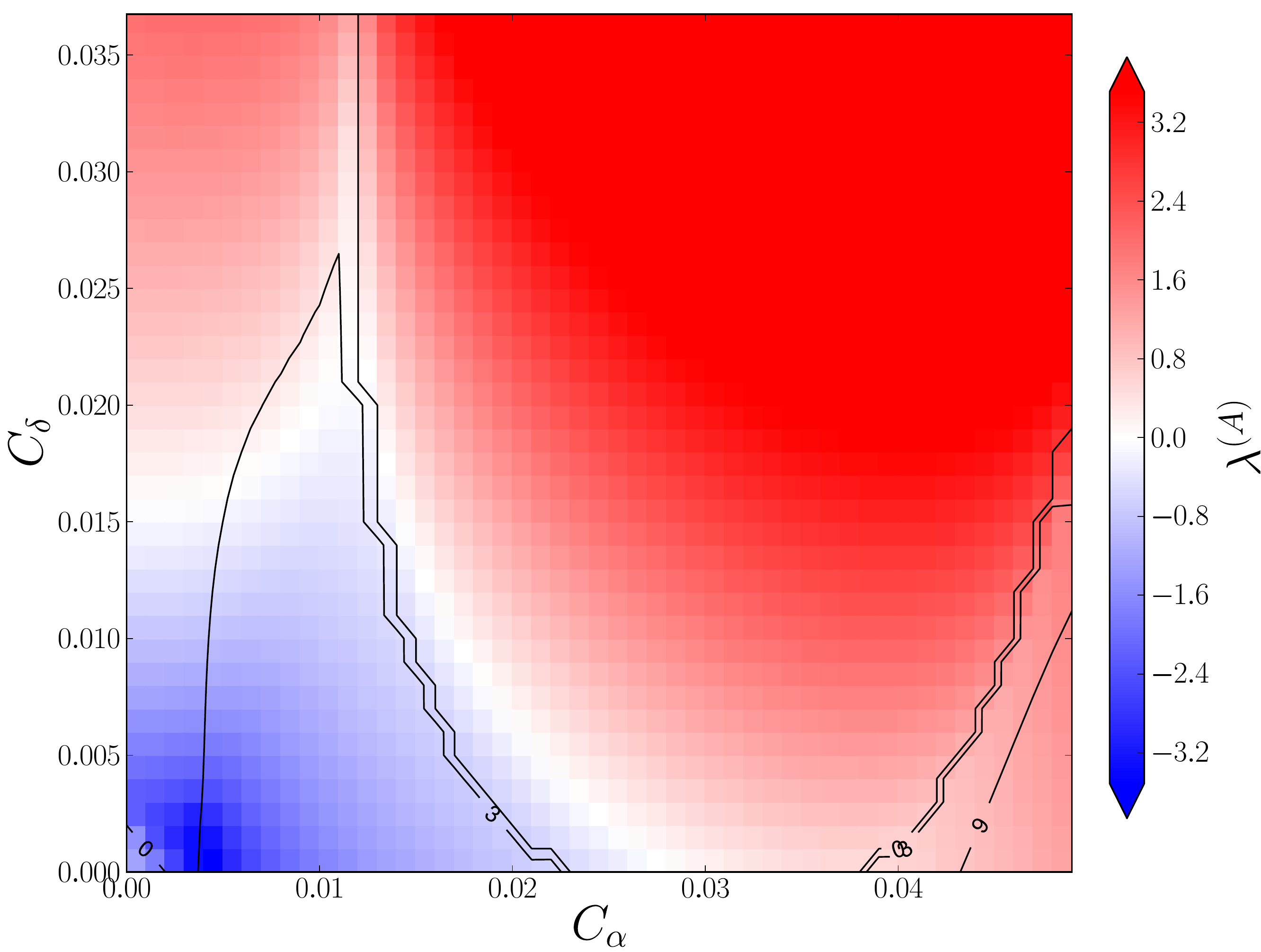}
\includegraphics[width=0.4\paperwidth,height=0.2\paperheight]{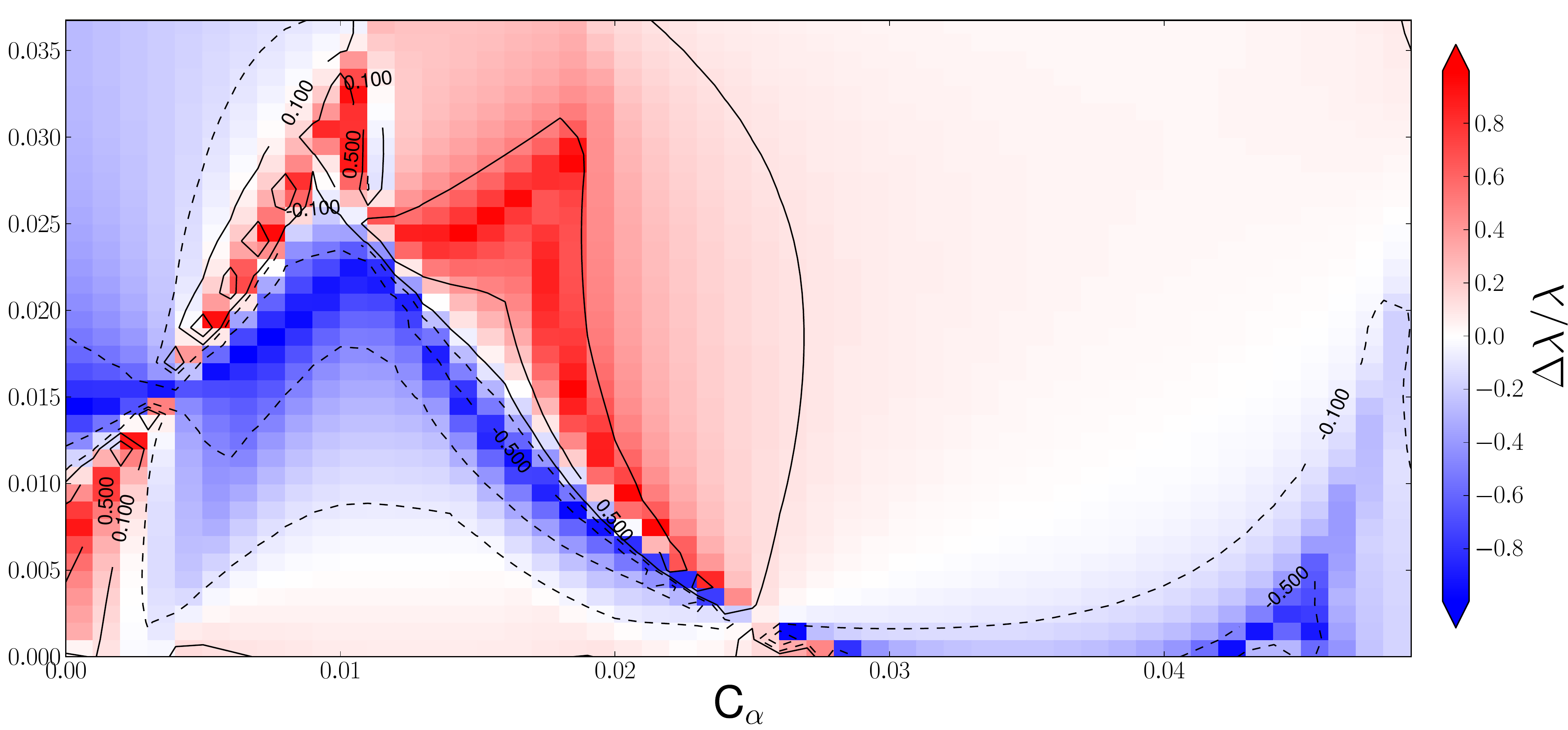}

\caption{\label{fig2}Left column shows the linear-stability diagrams for the
dynamo models with the meridional circulation speed $U_{0}=8,12$
and $16$ $\mathrm{m\cdot s^{-1}}$ from the top panel to bottom.
The color scale shows the growth rate of the most unstable A-mode,
$\lambda^{(A)}$. The contours show the oscillation frequency, $\omega^{(A)}$,
of this mode in the units of ${\displaystyle \frac{\eta_{T}^{(0)}}{R_{\odot}^{2}}}$.
The right column shows the relative growth rates of A- and S-modes:
${\displaystyle \frac{\Delta\lambda}{\lambda}=\frac{\left|\lambda^{(A)}\right|-\left|\lambda^{(S)}\right|}{\left|\lambda^{(A)}\right|+\left|\lambda^{(S)}\right|}}$,
where S-mode has the symmetric toroidal field relative to the equator.
The A-mode dominates in the regions colored in red. }
\end{figure}

\section{Results }

We calculate the dynamo solutions for a low level of the background
turbulent diffusivity, choosing $C_{\eta}=0.1$, in order to approximately match
the period of the eigenmodes with the solar cycle period. This corresponds
to the background diffusivity $\sim 10^{8}\mathrm{m^{2}s^{-1}}$. Figure
\ref{fig2} (left column) shows the linear-stability diagrams for
the dynamo models with the meridional circulation speed values equal
to $U_{0}=8,12$ and $16$ $\mathrm{m\, s^{-1}}$. The growth rate
$\lambda^{(A)}=Re\left(\sigma^{(A)}\right)$ of the first, most unstable
dynamo mode (A-mode) is shown by the color-scale plots in the $\left(C_{\alpha},C_{\delta}\right)$
plane, where $C_{\alpha}$ and $C_{\delta}$ are the free parameters
that control the strength of the $\alpha$ and $\Omega\times J$ effects.
We find that the dynamo instability region (represented by red color)
changes significantly with the increase of the meridional circulation
speed. For the slow meridional circulation it is found that the first
A-mode is stable and steady in the absence of the $\alpha$-effect
($C_{\alpha}=0$). It has the excitation threshold of $C_{\delta}\approx0.013$.
In the opposite limit, when $C_{\delta}=0$, the first mode is stable
and oscillating. Its oscillation frequency grows with the increase
of the $\alpha$-effect parameter $C_{\alpha}$. The excitation threshold
is $C_{\alpha}\approx0.025$ . The oscillation frequency of the mode
at the threshold is about ${\displaystyle 8\frac{\eta_{T}^{(0)}}{R_{\odot}^{2}}}$.

The right column in Figure \ref{fig2} shows the growth rate of the
first A-mode relative to the first S-mode. The relative difference
is characterized by parameter ${\displaystyle \frac{\left|\lambda^{(A)}\right|-\left|\lambda^{(S)}\right|}{\left|\lambda^{(A)}\right|+\left|\lambda^{(S)}\right|}}$.
This helps to identify the regions in the parameter space $\left(C_{\alpha},C_{\delta}\right)$
where the A-mode dominates the S-mode. We find that in the case of
$U_{0}=8$ $\mathrm{ms^{-1}}$ the A-mode is dominant for $C_{\alpha}\ge C_{\delta}$.
In this regime we look for a solar-type dynamo solution, because in
the solar dynamo the toroidal magnetic field is A-type (antisymmetric
relative to the equator).

\begin{figure}
\begin{centering}
\includegraphics[width=0.8\columnwidth]{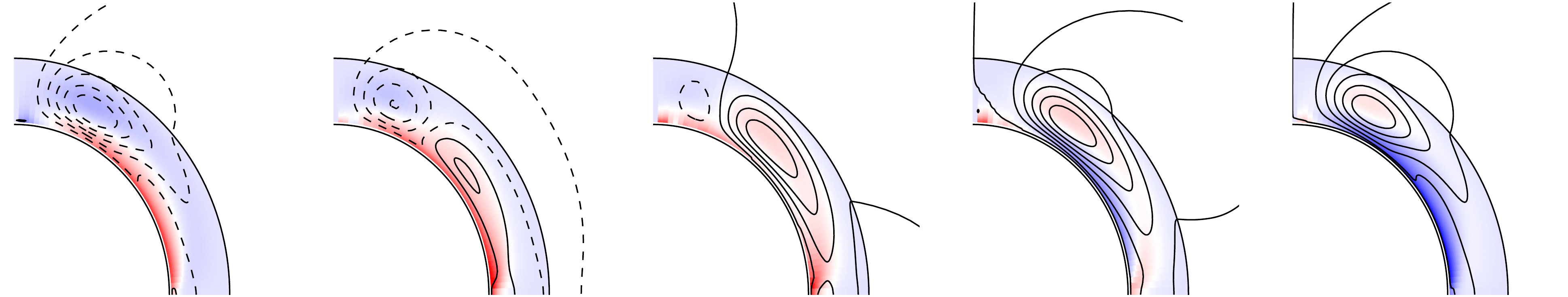} 
\par\end{centering}

\begin{centering}
\includegraphics[width=0.8\columnwidth]{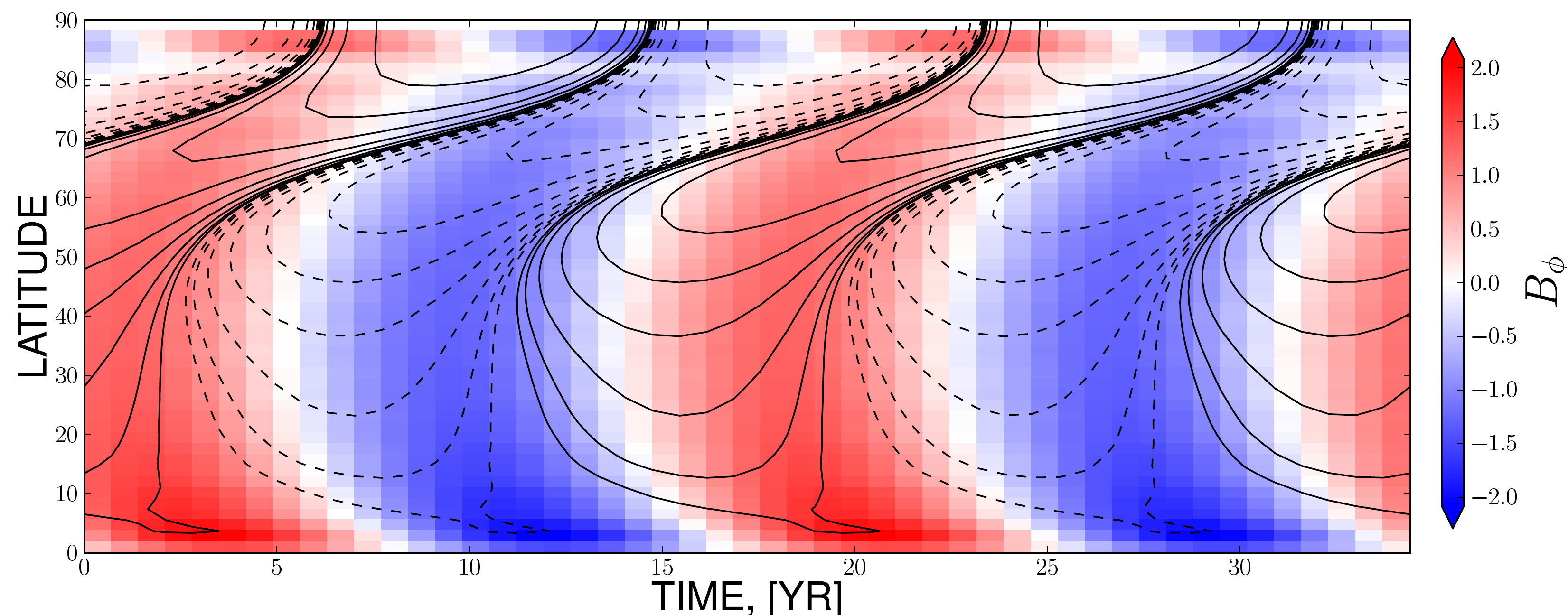} 
\par\end{centering}

\caption{\label{fig:MC8ms(a)} The snapshots for the magnetic
field variation inside convection zone (top) and the butterfly diagram
for this mode(bottom), for the meridional flow speed $U_{0}=8$ $\mathrm{ms^{-1}}$
and $C_{\alpha}=0.025$ and $C_{\delta}=0$.}
\end{figure}

For an example, we examine the case of $C_{\delta}=0$, $C_{\alpha}\approx0.025$
when the first A-mode has the frequency ${\displaystyle \approx8\frac{\eta_{T}^{(0)}}{R_{\odot}^{2}}}$.
Figure \ref{fig:MC8ms(a)} shows the snapshots
of the magnetic field variation inside the convection zone (top) and
the time-latitude {}``butterfly'' diagram for this mode (bottom).
The snapshots show that the toroidal magnetic field is concentrated
at the bottom of the convection zone, and the poloidal field is concentrated
in the polar regions. Also, the toroidal magnetic field is globally
distributed in the bulk of the convection zone. The maximum of the
toroidal field distribution drifts to the equator at the bottom of
the convection zone and moves to the pole near the surface. The bottom
panel of Figure \ref{fig:MC8ms(a)} shows the
butterfly diagrams of the toroidal field at the bottom of the convection
zone (color background) and for the radial magnetic fields at the
surface (contour lines). The toroidal magnetic field evolution pattern
has the polar and equatorial branches. The equatorial branch is strongly
concentrated to equator. The phase relation between the radial magnetic
field in the polar regions and the toroidal field in the equatorial
regions is in agreement with observations of the polar magnetic field
and the sunspot butterfly diagram, assuming that sunspots are
formed by emerging toroidal magnetic field. However, this dynamo mode lacks
the equatorial branch of the large-scale radial magnetic field, which
is also found in observations. The period of the dynamo period is
about 12 years. This is as twice as short compared to the solar magnetic
cycle. The period can be increased by further decreasing the diffusivity
parameter $C_{\eta}$ by a factor of 2 ($C_{\eta}\approx0.05$). However,
this leads to decrease of the excitation threshold and increase
of the effective magnetic Reynolds number ($R_{M}={\displaystyle \frac{U_{0}R_{\odot}}{\eta_{T}}}$).
This means that the S-mode becomes dominant, and solution no longer
corresponds to the solar dynamo. Figure \ref{fig2} shows that for
the case of $U_{0}=16$ $\mathrm{ms^{-1}}$ the S-mode dominates.

Inspecting the results in Figure 3 we can conclude that the increase
of the meridional circulation speed has two main effects on the dynamo
instability. Firstly, the larger $U_{0}$ the larger the unstable
area in the $\left(C_{\alpha},C_{\delta}\right)$ space, occupied
by the first unstable A-mode (associated with a non-oscillating dynamo
solution). Secondly, the S-mode becomes dominant near the instability
threshold everywhere, both for the case of $C_{\delta}=0$ and arbitrary
$C_{\alpha}$ and for the case of arbitrary $C_{\delta}$ and $C_{\alpha}=0$.
The combination of the $\alpha$ and $\Omega\times J$ effects can
make the A-mode dominant, but it represents a non-oscillating dynamo
solution.

\begin{figure}
\begin{centering}
\includegraphics[width=0.8\columnwidth]{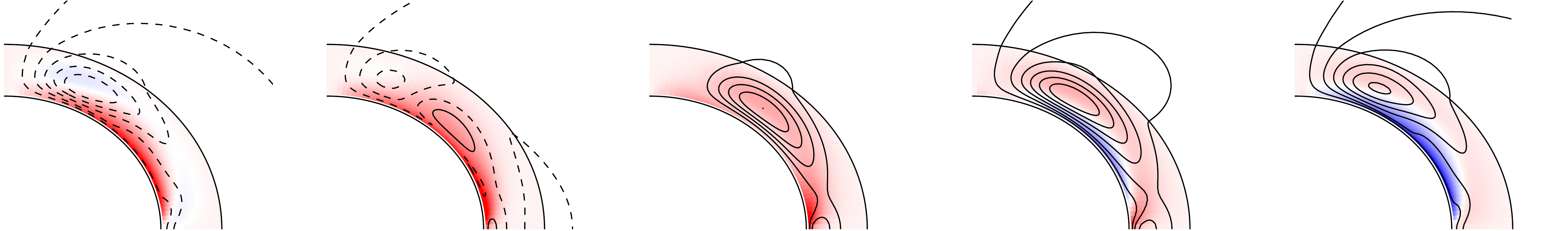} 
\par\end{centering}

\begin{centering}
\includegraphics[width=0.8\columnwidth]{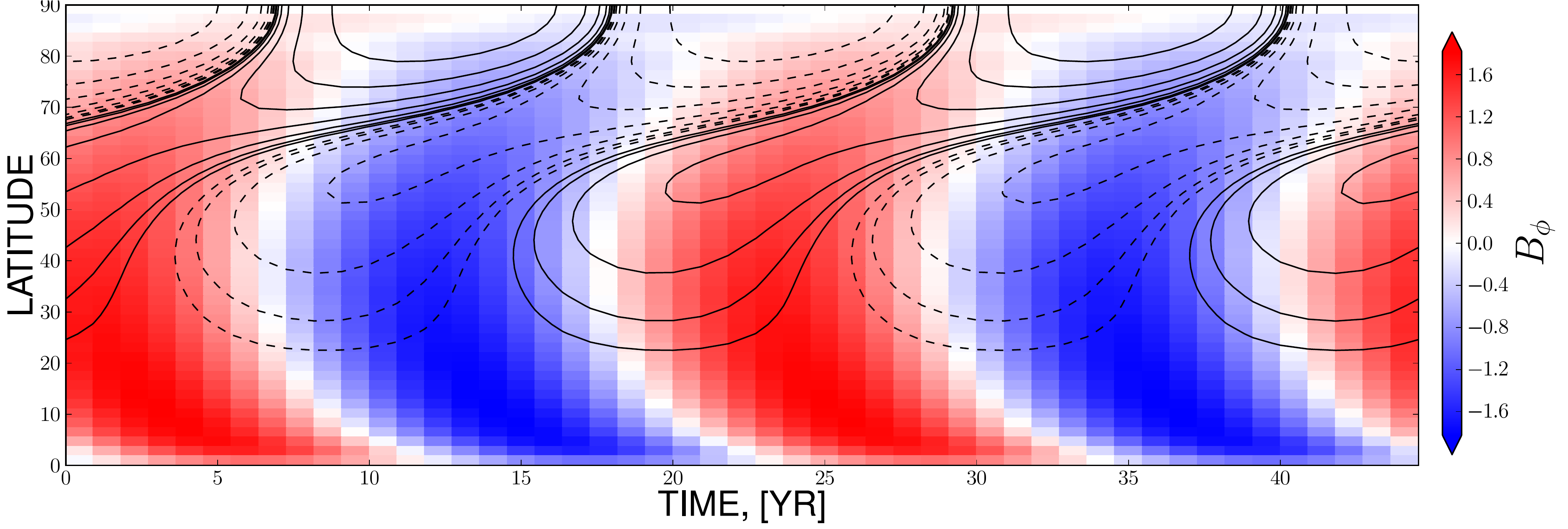} 
\par\end{centering}

\caption{\label{fig:MC8ms}The as in Figure \ref{fig:MC8ms(a)}
for a dynamo with $C_{\alpha}=C_{\delta}=0.015$}
\end{figure}

In another example, we examine the model, when the poloidal field
is generated both by the $\alpha$-effect and the $\Omega\times J$-effect,
e.g., $C_{\alpha}=C_{\delta}=0.015$ and $U_{0}=8$ $\mathrm{ms^{-1}}$
. The oscillation frequency of the first unstable A-mode is about
${\displaystyle 4\frac{\eta_{T}^{(0)}}{R_{\odot}^{2}}}$. Near the
excitation threshold, the A-mode is highly dominant over the first
S-mode. Figure \ref{fig:MC8ms} shows the snapshots
of the magnetic field variations inside the convection zone (top)
and the butterfly diagram for this mode(bottom). The snapshots of
the magnetic field evolution inside the convection zone are similar
to the previous case. However, the toroidal magnetic field is stronger
concentrated at the bottom of convection zone, and the polar branch
of the toroidal magnetic field evolution near is weaker near the surface.
The period of dynamo wave is about 24 years, close to the solar cycle.
Generally, we see a significantly better agreement with the observations
than in the previous case.

\begin{figure}
\begin{centering}
\includegraphics[width=0.45\textwidth]{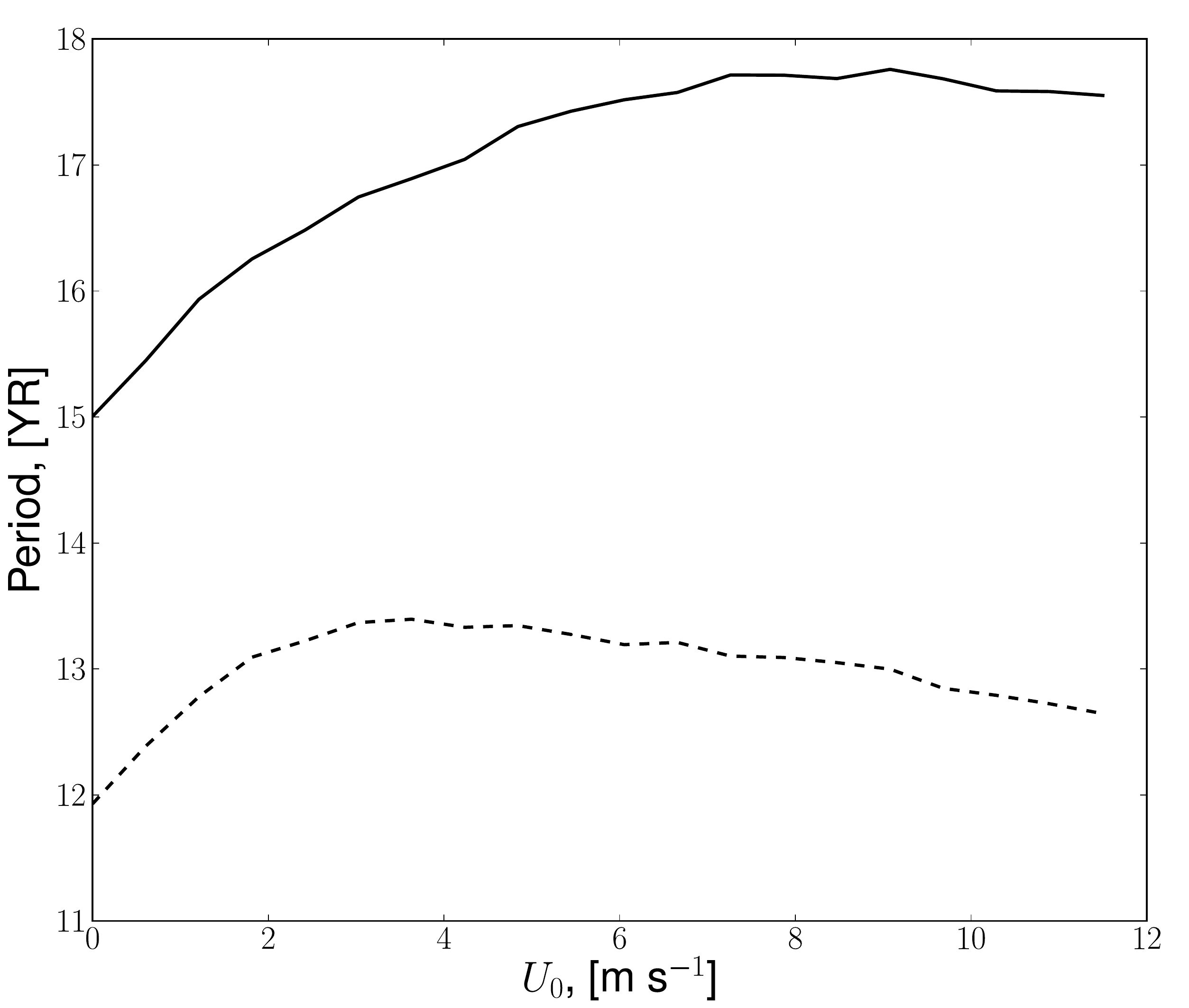}\includegraphics[width=0.45\textwidth]{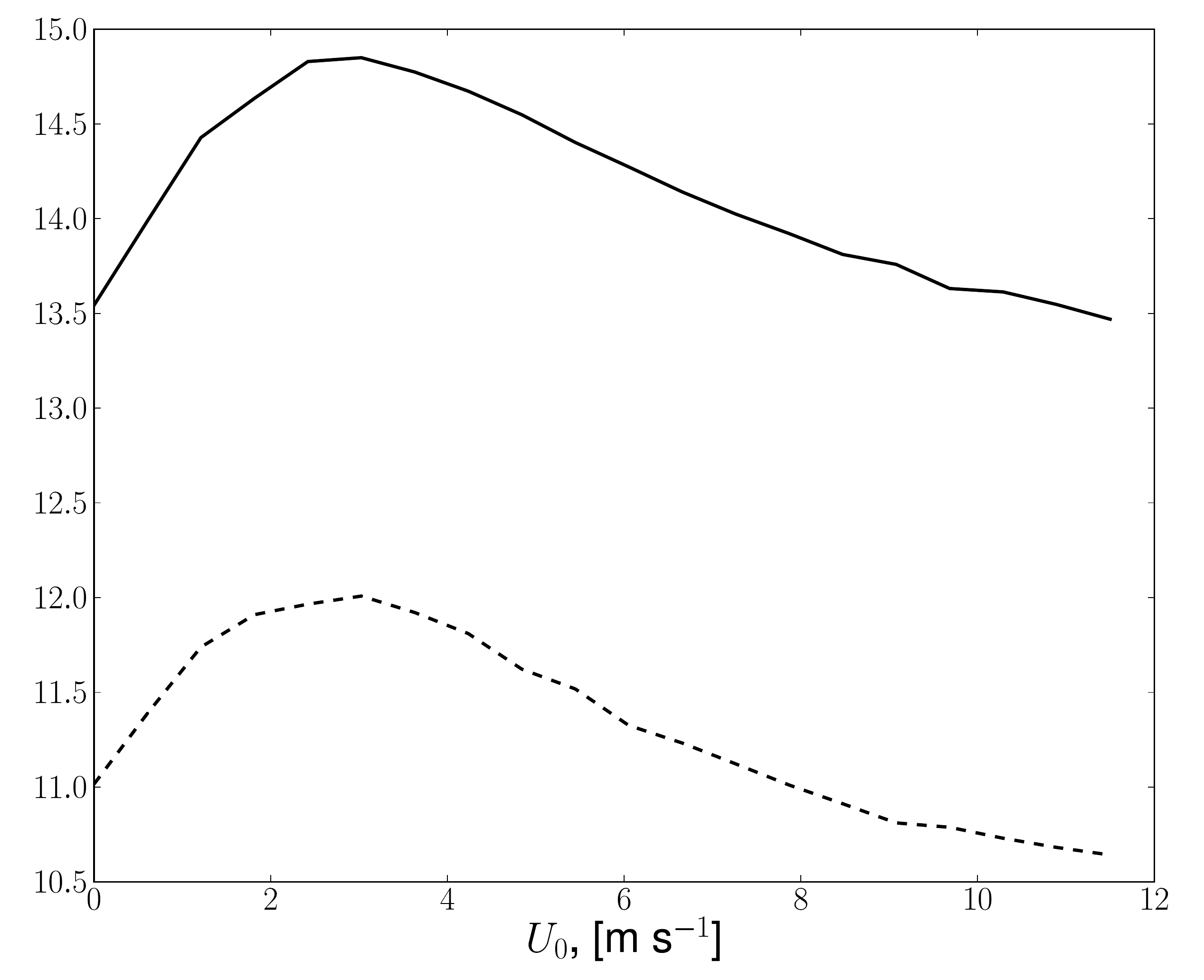} 
\par\end{centering}

\caption{\label{fig6}Dependence of the dynamo period on the meridional flow
speed $U_{0}$ along the stability boundary of the most unstable dipolar
mode for the dynamo models with the $\Omega\times J$ effect (solid
line) and without it (dashed line). Left, the results for complete
$\alpha$-effect and right for the $\alpha$-effect with $\Lambda^{(u)}=0$
(see, Eq.(\ref{eq:alpha}))}
\end{figure}

In Figure \ref{fig6} the dependence of the dynamo wave period on
the speed of the meridional flow along the stability threshold shown.
Contrary to previous results, e.g., \citep{2002A&A...390..673B,2009A&A...508....9S}
the period is not a monotonic function of the flow velocity. The main
reason is that here we use the meridional circulation with a different
depth dependence. We made a check how the presented results may depend
on the distribution of the alpha effect. For this,, we switched off
the effects of the turbulent mixing stratification, $\Lambda^{(u)}=0$
in Eq.(\ref{eq:alpha}). The dynamo period as a function of the meridional
flow speed for this case is shown in Figure \ref{fig6}(right). We
find that dependence of the dynamo period on the flow speed is much
stronger in case $\Lambda^{(u)}=0$ for both types of dynamo.

\section{Discussion and conclusions}

We have studied kinematic axisymmetric mean-field dynamo models for
a meridional circulation pattern with a deep seated stagnation point.
This kind of circulation is suggested by the mean-field models of
the angular momentum balance in the solar convection zone. We show
that by adjusting the turbulent sources of the poloidal magnetic field
generation and the turbulent diffusion strength it is possible to
construct a mean-field dynamo model that resembles in some aspects
the solar magnetic cycle. The most important features of the investigated
models are the following.



The maximum strength of the toroidal magnetic field, which is believed
to be responsible for the sunspots production, is concentrated near
the bottom of the convection zone. This field is transported to the
equatorial regions by the meridional flow. The meridional component
of the poloidal field is also concentrated at the bottom of the convection
zone. The large-scale radial field is concentrated near the poles.
It reverses sign when the maximum of the toroidal field gets close
to equator. This is not quite consistent with the solar observations,
which show that the polar field reversals happen earlier in the cycle.
Similar result was demonstrated in the kinematic flux-transport model
by \citet{2006ApJ...647..662R}. His model had a qualitatively similar
meridional circulation pattern, and the speed at the bottom of the
convection zone was as twice as slow compared to our case. We believe
that this feature (the phase relation) is inherent for this type of
the meridional flow, which produced conveyor-belt like circulation
of magnetic field \citep{2004ApJ...601.1136D}. The equatorward and
poleward conveyor bands are not well connected in our case because
circulation is quite weak in the bulk of convection zone.

We find that including the combined action of the $\alpha$ and $\Omega\times J$
effects for the poloidal magnetic field generation improve the agreement
of the model with observation. Contrary to the usual expectations
that come from the results of the flux-transport dynamo model we find
that the period of the dynamo cycle does not always become shorter when
the speed of the meridional circulation increases. In our model this
rule works for the amplitude of flow $>3\mathrm{m s^{-1}}$ in the case
of the $\alpha^{2}\Omega$
dynamo with the $\alpha$ effect dependent on the density stratification,
and for $>8\mathrm{m s^{-1}}$ in the case of the $\alpha^{2}\delta\Omega$ with the $\alpha$
effect dependent on both the density  and the turbulent diffusivity
stratifications. The dependence of the dynamo period
on the flow amplitude is much stronger if the alpha effect does not
depend on the turbulence intensity stratification, $\Lambda^{(u)}=0$.

Thus, by measuring the distributions of the magnetic activity and
meridional circulation characteristics on the Sun and possibly other
cool stars we may get indirect information about contribution of the
$\Omega\times J$ effect to the dynamo and the relative contributions
to the $\alpha$ effect due to density and the turbulent diffusivity
stratifications as well.

We conclude that the meridional flow pattern and speed have to be
considered among the most important constrains on the stellar dynamo.
Our results show the possibility of using helioseismic observations
of the meridional circulation for the diagnostic purpose of the solar
dynamo, because the dynamo properties significantly depends on the
depth of the flow stagnation point. 

\section{Acknowledgements}

This work was supported by NASA LWS NNX09AJ85G grant and partially
by RFBR grant 10-02-00148-a. 

\bibliographystyle{apj}


\section{Appendix}

Here we describe the components of the mean-electromotive force which
is used in the model. The tensor $\alpha_{i,j}$ represents the turbulent
alpha effect, and in accordance with P08 it is given by 
\begin{eqnarray}
\alpha_{ij} & = & \delta_{ij}\left\{ 3\eta_{T}\left(f_{10}^{(a)}\left(\mathbf{e}\cdot\boldsymbol{\Lambda}^{(\rho)}\right)+f_{11}^{(a)}\left(\mathbf{e}\cdot\boldsymbol{\Lambda}^{(u)}\right)\right)\right\} +\label{eq:alpha}\\
 & + & e_{i}e_{j}\left\{ 3\eta_{T}\left(f_{5}^{(a)}\left(\mathbf{e}\cdot\boldsymbol{\Lambda}^{(\rho)}\right)+f_{4}^{(a)}\left(\mathbf{e}\cdot\boldsymbol{\Lambda}^{(u)}\right)\right)\right\} \nonumber \\
 & + & 3\eta_{T}\left\{ \left(e_{i}\Lambda_{j}^{(\rho)}+e_{j}\Lambda_{i}^{(\rho)}\right)f_{6}^{(a)}+\left(e_{i}\Lambda_{j}^{(u)}+e_{j}\Lambda_{i}^{(u)}\right)f_{8}^{(a)}\right\} ,\nonumber 
\end{eqnarray}
 tensor $\gamma_{i,j}$ describes the turbulent pumping 
\begin{equation}
\gamma_{ij}=3\eta_{T}\left\{ f_{3}^{(a)}\Lambda_{n}^{(\rho)}+f_{1}^{(a)}\left(\mathbf{e}\cdot\boldsymbol{\Lambda}^{(\rho)}\right)e_{n}\right\} \varepsilon_{inj}-3\eta_{T}f_{1}^{(a)}e_{j}\varepsilon_{inm}e_{n}\Lambda_{m}^{(\rho)},\label{eq:pump}
\end{equation}
 and the $\eta_{ijk}$ term describes the anisotropic diffusion due
to the Coriolis force and the $\Omega\times J$ effect \citep{rad69},
\begin{equation}
\eta_{ijk}=3\eta_{T}\left\{ \left(2f_{1}^{(a)}-f_{1}^{(d)}\right)\varepsilon_{ijk}-2f_{1}^{(a)}e_{i}e_{n}\varepsilon_{njk}+f_{4}^{(d)}\delta_{ij}e_{k}\right\} ,\label{eq:diff}
\end{equation}
 functions $f_{\{1-11\}}^{(a,d)}$ (given below) depend on the Coriolis
number $\Omega^{*}=2\tau_{c}\Omega_{0}$ and the typical convective
turnover time in the mixing-length approximation is $\tau_{c}=\ell/u'$.
The turbulent diffusivity is parametrized in the form, $\eta_{T}=C_{\eta}\eta_{T}^{(0)}$,
where $\eta_{T}^{(0)}={\displaystyle \frac{u'\ell}{3}}$ is the characteristic
mixing-length turbulent diffusivity, $u'$ is the RMS convective velocity,
$\ell$ is the mixing length, $C_{\eta}$ is a constant to control
the intensity of turbulent mixing. The background turbulence is a
state of turbulent flows in the absence of the mean magnetic fields
and global rotation. The others quantities in Eqs.(\ref{eq:alpha},\ref{eq:pump},\ref{eq:diff})
are: $\mathbf{\boldsymbol{\Lambda}}^{(\rho)}=\boldsymbol{\nabla}\log\overline{\rho}$
is the density stratification scale, $\mathbf{\boldsymbol{\Lambda}}^{(u)}=\boldsymbol{\nabla}\log\left(\eta_{T}^{(0)}\right)$
is the scale of turbulent diffusivity, $\mathbf{e}=\boldsymbol{\Omega}/\left|\Omega\right|$
is a unit vector along the axis of rotation. Equations (\ref{eq:alpha},\ref{eq:pump},\ref{eq:diff})
take into account the influence of the fluctuating small-scale magnetic
fields, which can be present in the background turbulence (see discussions
in \citealp{pouquet-al:1975a,moff:78,vain-kit:83,1996A&A...307..293K,2005PhR...417....1B}).
In our paper, the parameter $\varepsilon={\displaystyle \frac{\overline{\mathbf{b}^{2}}}{\mu_{0}\overline{\rho}\overline{\mathbf{u}^{2}}}}$,
which measures the ratio between the magnetic and kinetic energies
of fluctuations in the background turbulence, is assumed equal to
1. This corresponds to the energy equipartition.

Below we give the functions of the Coriolis number defining the dependence
of the turbulent transport generation and diffusivities on the angular
velocity: 
\begin{eqnarray*}
f_{1}^{(a)} & = & \frac{1}{4\Omega^{*\,2}}\left(\left(\Omega^{*\,2}+3\right)\frac{\arctan\Omega^{*}}{\Omega^{*}}-3\right),\\
f_{3}^{(a)} & = & \frac{1}{4\Omega^{*\,2}}\left(\left(\left(\varepsilon-1\right)\Omega^{*\,2}+\varepsilon-3\right)\frac{\arctan\Omega^{*}}{\Omega^{*}}+3-\varepsilon\right),\\
f_{4}^{(a)} & = & \frac{1}{6\Omega^{*\,3}}\left(3\left(\Omega^{*4}+6\varepsilon\Omega^{*2}+10\varepsilon-5\right)\frac{\arctan\Omega^{*}}{\Omega^{*}}-\left((8\varepsilon+5)\Omega^{*2}+30\varepsilon-15\right)\right),\\
f_{5}^{(a)} & = & \frac{1}{3\Omega^{*\,3}}\left(3\left(\Omega^{*4}+3\varepsilon\Omega^{*2}+5(\varepsilon-1)\right)\frac{\arctan\Omega^{*}}{\Omega^{*}}-\left((4\varepsilon+5)\Omega^{*2}+15(\varepsilon-1)\right)\right),\\
f_{6}^{(a)} & = & -\frac{1}{48\Omega^{*\,3}}\left(3\left(\left(3\varepsilon-11\right)\Omega^{*2}+5\varepsilon-21\right)\frac{\arctan\Omega^{*}}{\Omega^{*}}-\left(4\left(\varepsilon-3\right)\Omega^{*2}+15\varepsilon-63\right)\right),\\
f_{8}^{(a)} & = & -\frac{1}{12\Omega^{*\,3}}\left(3\left(\left(3\varepsilon+1\right)\Omega^{*2}+4\varepsilon-2\right)\frac{\arctan\Omega^{*}}{\Omega^{*}}-\left(5\left(\varepsilon+1\right)\Omega^{*2}+12\varepsilon-6\right)\right),\\
f_{10}^{(a)} & = & -\frac{1}{3\Omega^{*\,3}}\left(3\left(\Omega^{*2}+1\right)\left(\Omega^{*2}+\varepsilon-1\right)\frac{\arctan\Omega^{*}}{\Omega^{*}}-\left(\left(2\varepsilon+1\right)\Omega^{*2}+3\varepsilon-3\right)\right),\\
f_{11}^{(a)} & = & -\frac{1}{6\Omega^{*\,3}}\left(3\left(\Omega^{*2}+1\right)\left(\Omega^{*2}+2\varepsilon-1\right)\frac{\arctan\Omega^{*}}{\Omega^{*}}-\left(\left(4\varepsilon+1\right)\Omega^{*2}+6\varepsilon-3\right)\right).\\
f_{1}^{(d)} & = & \frac{1}{2\Omega^{*\,3}}\left(\left(\varepsilon+1\right)\Omega^{*\,2}+3\varepsilon-\left(\left(2\varepsilon+1\right)\Omega^{*\,2}+3\varepsilon\right)\frac{\arctan\left(\Omega^{*}\right)}{\Omega^{*}}\right),\\
f_{4}^{(d)} & = & \frac{1}{2\Omega^{*\,3}}\left(\left(2\Omega^{*\,2}+3\right)-3\left(\Omega^{*\,2}+1\right)\frac{\arctan\left(\Omega^{*}\right)}{\Omega^{*}}\right).
\end{eqnarray*}

\end{document}